\documentclass[fleqn,usenatbib]{mnras}

\usepackage{newtxtext,newtxmath}
\usepackage[T1]{fontenc}

\DeclareRobustCommand{\VAN}[3]{#2}
\let\VANthebibliography\thebibliography
\def\thebibliography{\DeclareRobustCommand{\VAN}[3]{##3}\VANthebibliography}

\usepackage{graphicx}	% Including figure files
\usepackage{amsmath}	% Advanced maths commands

\usepackage{svg}

\newcommand{\microHz}{\,$\mu$Hz} % microHz
\title[Solar acoustic power variation]{Unexpected solar-cycle variation of acoustic mode power in Sun-as-a-star observations}

\author[R. Howe et al.]{
Rachel Howe,$^{1,2}$\thanks{E-mail: r.howe@bham.ac.uk (RH)}
W. J. Chaplin,$^{1,2}$
Y. P. Elsworth,$^{1,2}$
S. J. Hale,$^{1,2}$
and M. B. Nielsen$^{1,2,3}$
\\
$^{1}$School of Physics and Astronomy, University of Birmingham, Edgbaston, Birmingham B15 2TT, UK\\
$^{2}$Stellar Astrophysics Centre (SAC), Department of Physics and Astronomy, Aarhus University, Ny Munkegade 120, DK-8000 Aarhus C, Denmark\\
$^{3}$Center for Space Science, NYUAD Institute, New York University Abu Dhabi, PO Box 129188, Abu Dhabi, United Arab Emirates
}

\date{Accepted XXX. Received YYY; in original form ZZZ}

\pubyear{2022}

\begin{document}
\label{firstpage}
\pagerange{\pageref{firstpage}--\pageref{lastpage}}
\maketitle

\begin{abstract}
We examine the solar-cycle variation of the power in the low-degree helioseismic modes by looking at binned power spectra from 45 years of observations with the Birmingham Solar Oscillations Network, which provides a more robust estimate of the mode power than that obtained by peak fitting. The solar-cycle variation of acoustic mode power in the five-minute band is clearly seen. Unusually, even though Cycle 24 was substantially weaker in terms of surface magnetic activity than Cycle 23, the reduction in mode power at solar maximum is very similar for the two cycles, suggesting that the relationship between mode power and magnetic activity is more complex than has previously been thought. This is in contrast to the mode frequencies, which show a strong correlation with activity with only subtle differences between in the response across different solar cycles.
%No references should appear in the abstract.
\end{abstract}


%%%%%%%%%%%%%%%%%%%%%%%%%%%%%%%%%%%%%%%%%%%%%%%%%%

%%%%%%%%%%%%%%%%% BODY OF PAPER %%%%%%%%%%%%%%%%%%

\section{Introduction}

While the main tool of helioseismology, the study of the Sun's modes of oscillation, is to use frequencies and rotational splitting coefficients of the modes (mainly the so-called $p$-modes) to infer the Sun's internal structure and dynamics, the other properties of the modes, such as amplitude, lifetime, and asymmetry, are sensitive to conditions in and around the photospheric layer that vary with the level of magnetic activity over the solar cycle. In this work we revisit the variation of the mode power in low-degree modes using integrated-light observations from the  Birmingham Solar Oscillations (BiSON) network \citep[][]{2016SoPh..291....1H} that stretch back over four solar cycles. It is well established that acoustic power in the five-minute band is suppressed in active regions where the magnetic field is strong \citep[e.g.][]{1998ApJ...504.1029H}. On global scales and timescales of years, this sensitivity to magnetic-field strength translates to a solar-cycle variation in the mode amplitude.

Before discussing solar-cycle changes in the strength of the oscillation modes, it is important to be precise about the definitions of the words ``power,'' ``height,'' and ``amplitude'' in this context. The profile of an oscillation mode in the acoustic power spectrum (derived from the Fourier transform of a time series of observations) is generally represented by a Lorentzian function, which in modern analysis may be modified by an asymmetry term. If such a peak has height $h$ and width $\Gamma$, the integrated area below it, sometimes referred to as the power ($P$), is $\pi \Gamma h$ and the quantity we call the amplitude $A$ is the square root of this.

The anticorrelation of mode power with activity was reported by \cite{1993MNRAS.265..888E} in early BiSON data covering the period 1981\,--\,1992, from the maximum of Solar Cycle 21 to just after that of Cycle 22. They reported an increase of $35\pm5$ per cent in the mode power from solar maximum to solar minimum for modes of low angular degree, based on estimating the mode peak parameters using least-squares fitting to the spectrum and applying ad-hoc corrections for the effects of the varying and mostly low duty cycle. These corrections are required because when the time series of observations has gaps, the observed mode power is redistributed away from the central peak as the solar power spectrum is convolved with that of the duty cycle; this results in broader, less prominent mode peaks and increased background power between the peaks, as well as the ``sidelobe'' peaks separated in frequency from the true peak by 1/day or {11.57\microHz} that occur when there are regular daily interruptions. The sidelobe spacing is close to the separation between pairs of modes separated by one radial order ($n$) and two degrees ($l$) in the low-degree $p$-mode spectrum.

\cite{2000MNRAS.313...32C} looked at a slightly later epoch, 1991\,--\,1996, of BiSON data, when the duty cycle was much better, and used a maximum-likelihood (MLE) peak-fitting technique. They reported a $46\pm 5$ per cent decrease in the mode heights, and a mean decrease of $22\pm3$ per cent in the power along with a $24\pm 5$ per cent increase in the mode linewidth, between solar maximum and solar minimum. Together with the lack of detectable change in the rate of energy being supplied to the mode, they interpreted this relationship between power and linewidth changes as evidence that the change in power was driven by the damping of the modes rather than their excitation. 

\cite{2003ApJ...588.1204H} looked at frequency, mode width, and mode height variations with magnetic activity  over the period from September 1995 to June 2000, using BiSON data alongside disk-integrated data from the resolved-Sun instruments of the Global Oscillations Network Group (GONG). They reported a decrease in mode height of about 2.5 per cent for each 1 Gauss increase in the integrated magnetic-field strength, accompanied by an increase in the mode width about half as large, which would correspond to a power decrease of about 1.25 per cent per Gauss. For comparison, this magnetic index \citep[described in detail by ][]{2017MNRAS.464.4777H} changed by $\approx$ 15\,G between the maximum of Cycle 22 and the subsequent minimum, which would correspond to about a 20 per cent change in the power, consistent with the \cite{2000MNRAS.313...32C} result. This result seems to reflect a weaker activity dependence than that reported by \cite{1993MNRAS.265..888E}, suggesting that interpreting the results may not be completely straightforward. These early studies used data with a time span of less than a single solar cycle.

More recently, \cite{2015MNRAS.454.4120H} looked at amplitude, width, and frequency variations in the BiSON data from 1992\,--\,2014 and compared the results with those from fits to synthetic ``SolarFLAG'' data designed to simulate the properties of BiSON data. They used both MLE fitting and a Bayesian Markov-chain Monte-Carlo (MCMC) technique and obtained similar results for each. The BiSON amplitude variation between the highest and lowest activity level (as measured from the 10.7\,cm radio flux) in that analysis was found to be $\approx$ 8\,--\,9 per cent, corresponding to a power change of 16\,--\,18 per cent. They concluded that this result was in reasonable agreement with the \cite{2000MNRAS.313...32C} result given that the 2015 work analysed longer time series and hence there were smaller differences in average activity level between spectra than would be seen with shorter ones. Comparison with the SolarFLAG results, where the data were constructed with a known linear relationship between amplitude and activity, suggested that the measured change was underestimated by about 10 per cent; this underestimate may also affect the work of \cite{2000MNRAS.313...32C}, who used a similar analysis. Neither the \cite{2003ApJ...588.1204H} nor the \cite{2015MNRAS.454.4120H} analysis attempted to correct for the differential effects of duty cycle, as the duty cycle was relatively flat throughout the period analysed. \cite{2015MNRAS.454.4120H} also noted that the decrease in amplitude for the BiSON data in Solar Cycle 24 appeared to be slightly larger than expected given the weaker activity level. This was also remarked on by \cite{2015AdSpR..56.2706B}, who were also analyzing amplitude and linewidth changes in BiSON data, and it raises questions as to whether it is appropriate to consider the amplitude change as a fixed function of activity level.

The fractional change in the mode amplitude, power, and linewidth with activity depends on frequency, being strongest at the peak of the five-minute power and weakest at the extremes. This frequency dependence was marginally detected in BiSON data by \cite{2015MNRAS.454.4120H}; it is more clearly seen in resolved-Sun observations, for example by \cite{2001ApJ...563..410R, 2004ApJ...608..562H} using the ring-diagram technique of local helioseismology. The local helioseismic analysis also reveals that around (or a little below) the acoustic cut-off frequency the trend reverses, with higher amplitudes being seen for higher-frequency modes or so-called ``pseudomodes'' in or around active regions than in quiet Sun. 

All of the work referenced above looked at the mode parameters as inferred from fitting to the $p$-mode spectrum. 
In this work, we look at the solar-cycle power variation of the modes observed by BiSON in a simpler and potentially more robust way by considering instead all the power in a given frequency band. This approach mitigates the issue of power being redistributed away from its real location in frequency by the window function.

\section{Data and analysis}
\subsection{Observational data}
\begin{figure}
\centering
%\includesvg[width=\linewidth]{fig1.svg}%{fillrf.svg}
\includegraphics[width=\linewidth]{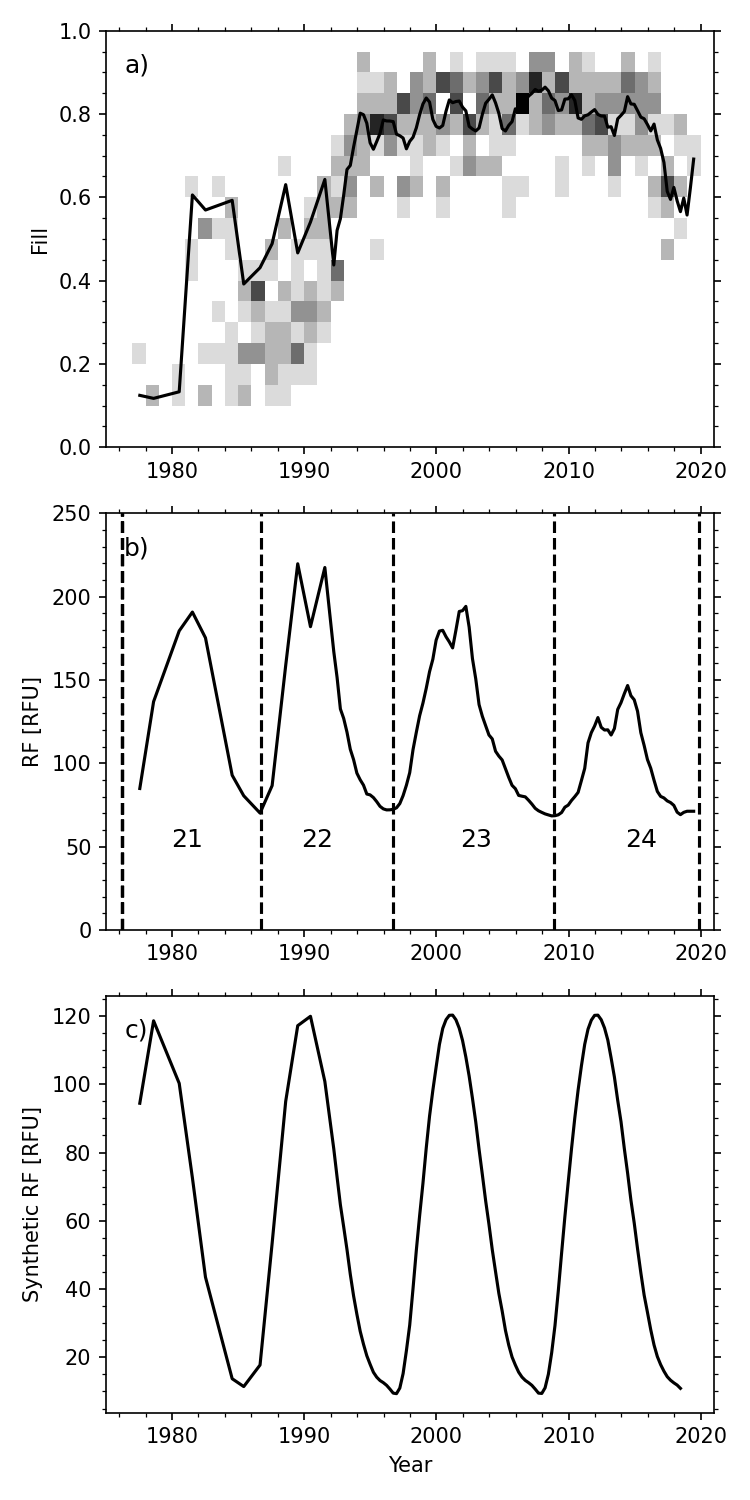}
\caption{a): Duty cycle of the BiSON network. The solid curve represents the fill for the spectra analysed here, and the background shading shows a 2d histogram of the fill in 30-day intervals binned by year. Intervals with duty cycle less than 10 per cent are excluded. b): Observed RF index for the observation periods used in the analysis. The dashed vertical lines indicate the solar minima and the solar cycle numbers are shown.
c): Synthetic activity proxy used for the SolarFLAG data.}
\label{fig:fig1}
\end{figure}

We use Sun-as-a-star radial-velocity observations from the BIrmingham Solar-Oscillations Network (BiSON), a worldwide network of ground-based observing stations making Doppler measurements of the integrated solar surface velocity in one of the  Potassium D lines. The full BiSON dataset begins in 1975 and continues up to the present, but initially coverage was sparse and intermittent, with only a few weeks per year of single- or dual-site observations, prior to the deployment of the modern network in the early 1990s \citep{1996SoPh..168....1C}.   From around 1994 to 2016 the six-station network achieved a duty cycle of over 75 per cent \citep{2016SoPh..291....1H}, 
and the duty cycle has remained above 60 per cent except for a month-long interruption in the early stages of the COVID-19 pandemic in March--April 2020. The duty cycle (``fill'', the fraction of the total time in each interval for which there are observations) for the BiSON data analysed here is shown in Figure~\ref{fig:fig1}(a), which also shows a two-dimensional histogram of the fill for 30-day periods for each year.

The BiSON time series was prepared as described by \cite{doi:10.1093/mnras/stu803} and then divided into a series of segments so that the temporal variation of the Fourier power spectrum could be studied. 
Because of the sparseness of the data prior to the deployment of the full network, for the years 1977\,--\,1991 we choose the best 64-day period for each calendar year, this interval being chosen as a convenient number similar to the length of a typical observing campaign. The years 1975, 1976, and 1978, where there was no 64-day period with a duty cycle greater than 10 per cent, were excluded from the analysis. There is an intractable problem with the calibration of the BiSON data for 1983, and this year is therefore also omitted from the analysis. For the data from 1992 onward we use overlapping 365-day periods with start dates spaced at 91.25 days.  A Fourier power spectrum was computed for each period. The spectra were scaled to satisfy Parseval's theorem such that the units of power are (m\,s$^{-1}$)$^2$ per Hertz, corrected for the zero-filled missing data by dividing the power by the fill
$f$. 

The 40-second cadence of the BiSON network gives a Nyquist frequency of 12.5 mHz. We divide this frequency range into  92 equally spaced bins approximately {135\,\microHz} wide and average the power across each bin; this spacing corresponds to that between modes of the same angular degree and adjacent radial order over most of the $p$-mode spectrum. Figure~\ref{fig:powspec} shows all of the binned BiSON spectra plotted on the same linear--log axes. While the peak of power in the five-minute band is always clear, there are considerable variations in the background levels at low and high frequencies, reflecting the varying performance of the network over time; the earliest data had both poor duty cycle and higher instrumental noise levels.

\subsection{Quantifying solar activity}

In order to look for solar-cycle variations in any feature of the oscillation spectrum, we need a quantitative measure of the activity level over time, also known as a solar-activity proxy.
In this work we are concerned only with ``Sun-as-a-star'' or integrated-light observations, so it is appropriate to use a solar-activity proxy that is integrated over the solar disk. Although earlier work has used different proxies, as discussed above, here we concentrate on the RF index or 10.7\,cm radio flux \citep[RF; ][]{2013SpWea..11..394T}. The values corresponding to the BiSON time series analysed in this work are shown in Figure~\ref{fig:fig1}(b). The daily RF index was averaged over the period covered by each spectrum, with the average weighted by the daily BiSON duty cycle; the difference between this weighted average and an unweighted one over the whole observing period is only important in the early years when the data were sparse.

\subsection{Artificial data}

To validate our results, we use synthetic data based on the ``SolarFLAG'' dataset that was also used (and described in detail) by \cite{2015MNRAS.454.4120H}. The basic SolarFLAG time series was designed to correspond to BiSON observations between 1996 and 2007. For the previous study we had generated a number of 11-year time series with the same synthetic activity cycle but independent realizations of the random noise; for this work we concatenated four such series to create a 44-year, four-cycle dataset with the same duty cycle as the BiSON observations. Figure~\ref{fig:fig1}(c) shows the 
synthetic quantity that emulates the solar-cycle RF variation in the SolarFLAG data. Note that in the SolarFLAG data, each cycle has the same activity pattern and strength, which is not the case for the real observations; also, although the synthetic proxy has a similar scale of variation to a real solar cycle, it goes to zero at solar minimum, while the real index never goes below about 70 flux units (RFU). The relationship between activity and mode amplitude change at a given frequency is assumed to be a fixed, linear one, which may not be realistic. Furthermore, the artificial data do not replicate the variations in instrumental noise levels as the  network evolved.

\begin{figure}
    \centering
    %\includesvg[width=\linewidth]{fig2.svg}%{bisonlines.svg}
    \includegraphics[width=\linewidth]{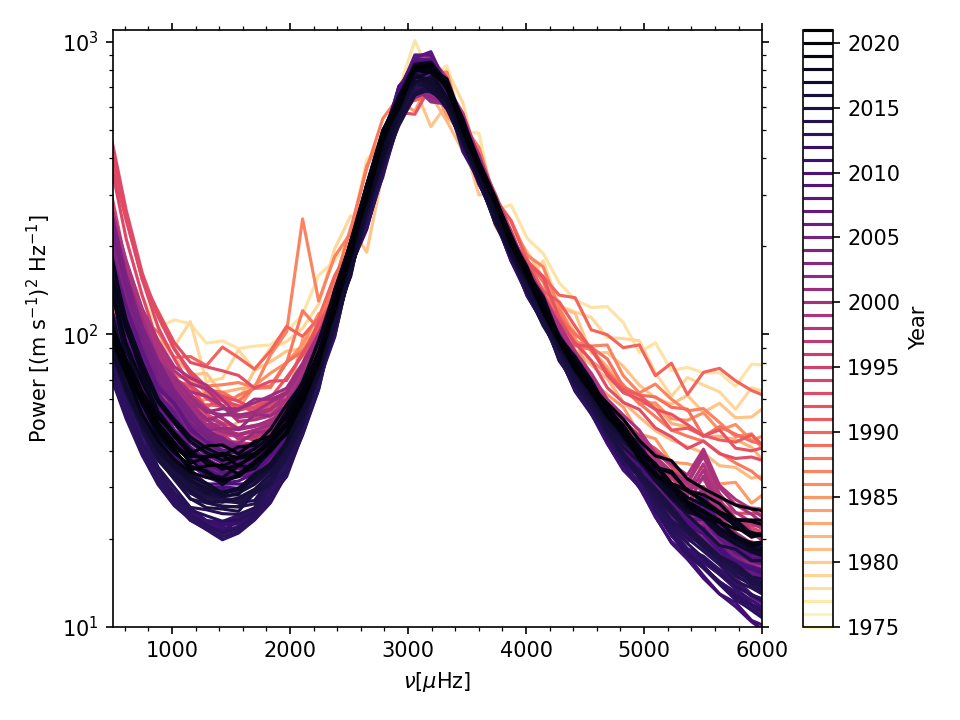}
    \caption{The BiSON spectra analysed in this work, averaged in {135-\microHz} bins and colour-coded by year.}
    \label{fig:powspec}
\end{figure}

\section{Results}

\subsection{Power variation from the power spectra}
We use Solar Cycle 23, for which we have the best coverage, as a reference, and we look at the fractional change of power in a given frequency band relative to the power $P_{23}(\nu)$ averaged over that cycle, $\delta P/P_{23} \equiv [P(\nu,t)-P_{23}(\nu)]/P_{23}(\nu)$, where $\nu$ denotes a frequency range and $t$ is time.

For each of the 11 {135\,\microHz} bins covering the range from 2445 to {3940 \microHz}, that is, the main peak of the five-minute oscillations, 
we calculate the mean power in the bin for each spectrum and the fractional shift in this power relative to the average over all the samples for that bin, and then we average these fractional shifts over the 11 bins to obtain a ``mean power-shift'' value for each spectrum.

\begin{figure*}
    \centering
    %\includesvg[width=\linewidth]{fig3.svg}%{fig3_2022-04-27.svg}
\includegraphics[width=\linewidth]{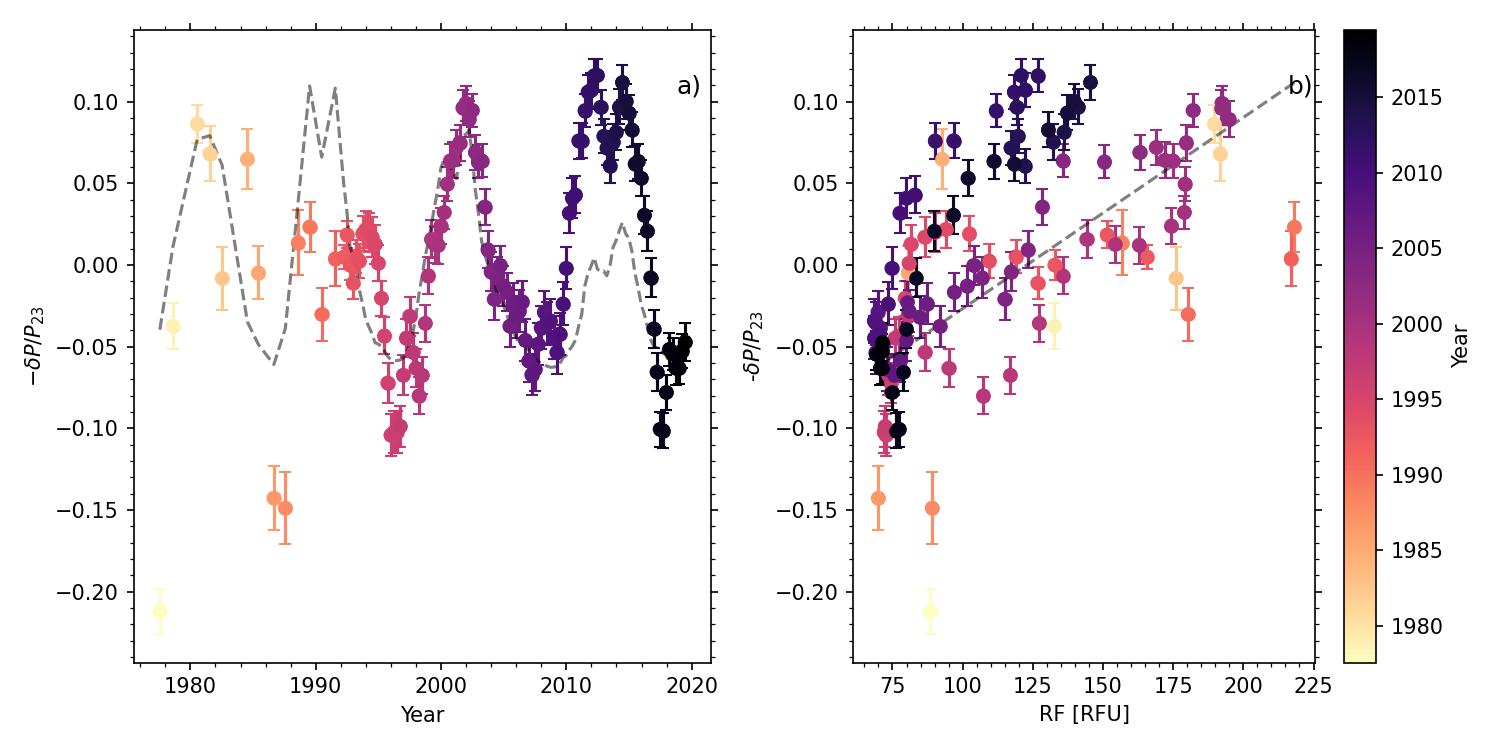}
    \caption{Sign-reversed fractional power change for BiSON data averaged over the 11 frequency bands covering the range from 2445 to 3940\,{\microHz} relative to the mean over Solar Cycle 23. Spectra are from 64-day time series for dates before 1992 and 365-day series thereafter, (left) as a function of time and (right) as a function of RF with the points colour-coded by year. The dashed lines in each panel represent a linear fit to the Cycle 23 RF, with slope 0.15 per cent per RFU.} 
    \label{fig:temporal}
\end{figure*}

Figure~\ref{fig:temporal} shows the 
sign-reversed average $\delta P/P_{23}$ for the five-minute band
plotted as a function of time and of the activity index, RF. 

One way of describing the variation that has been used in most previous work is to say that after appropriate corrections the fractional power change in a given frequency band can be expressed (roughly) as a linear function of the activity index:
\begin{equation}
{\frac{P(\nu)-P_{23}(\nu)}{P_{23}(\nu)}}={a_{RF}*RF+c},
\label{eq:eq1}
\end{equation}
where $a_{RF}$ is a ``sensitivity coefficient'' and $c$ is a constant.

It is clear in Figure
%s ~\ref{fig:wjcfig} and 
 \ref{fig:temporal}(a) that the lowest level of the power at the maximum of  Cycle 24 is similar to that at the maximum of  Cycle 23 even though the activity was significantly lower at the Cycle 24 maximum. This means that the dependence of power on activity in Cycle 24 appears steeper than that in Cycle 23.  In Figure~\ref{fig:temporal}(b) we also plot the trend derived using a linear least-squares fit to Equation~\ref{eq:eq1} for Solar Cycle 23 (dates between 1996 and 2009). We can see clustering of the darker-coloured points for Cycle 24 above the Cycle 23 trend-line. As the dependence for the two intermediate-sized Cycles 21 and 22 seems to lie between the Cycle 23 and Cycle 24 extremes, one could almost say that the sensitivity of the modes to activity level is inversely proportional to the strength of the cycle, but due to the sparser data in the earlier cycles it is hard to make a firm statement about this. We note that the extent of the frequency band used was chosen based on the appearance of the spectra in Figure~\ref{fig:powspec}, but the result holds even if we vary the boundaries by up to {500\,\microHz} at either extreme. The strongest (and most strongly varying) modes in the centre of the five-minute band dominate the average. 

\begin{figure*}
    \centering
    %\includesvg[width=\linewidth]{fig4.svg}%{fig4_2022-04-27.svg}
\includegraphics[width=\linewidth]{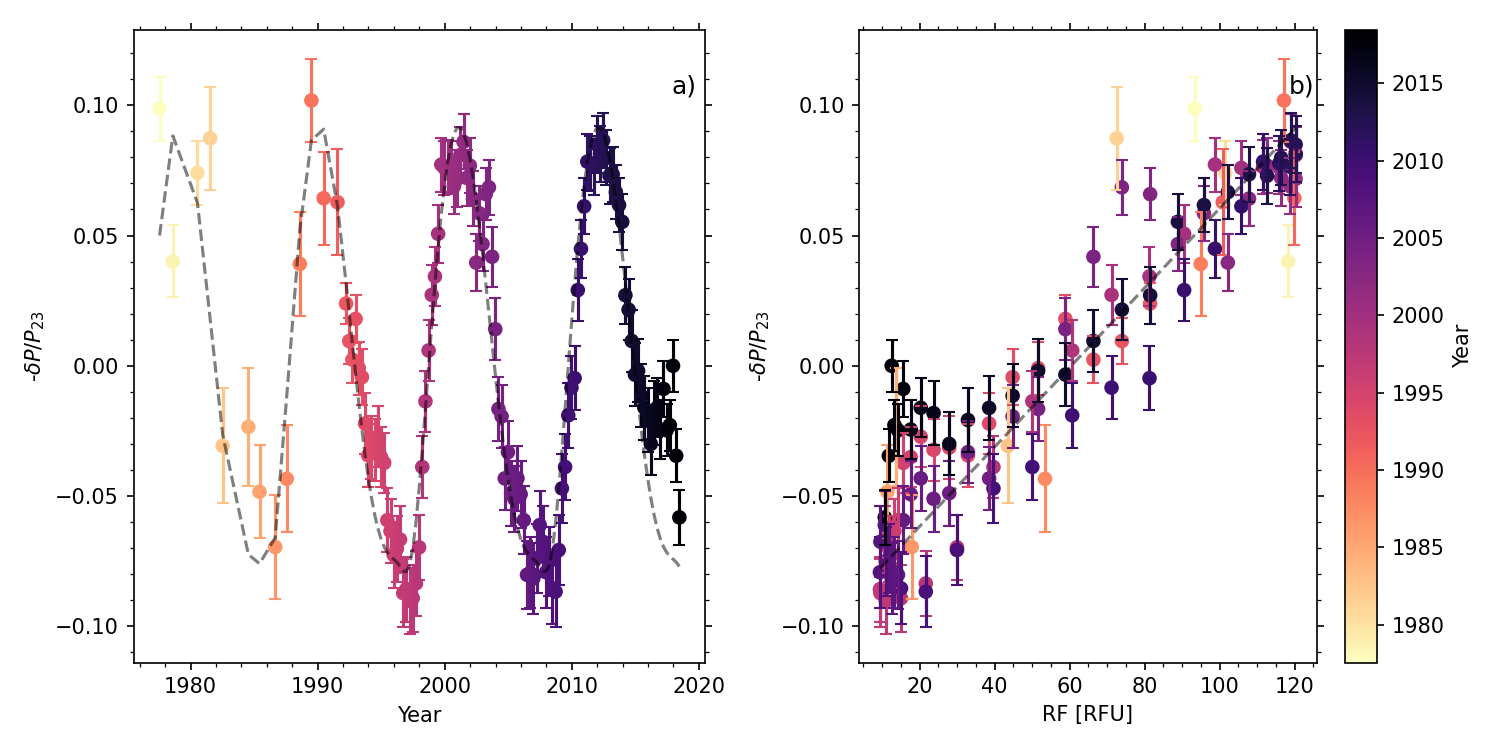}
    \caption{For SolarFLAG synthetic data with the BiSON duty cycle, sign-reversed fractional power change in the band from 2445 to 3940\,{\microHz} relative to the mean over Solar Cycle 23. 
    Spectra are from 64-day time series for dates before 1992 and 365-day series thereafter, (left) as a function of time and (right) as a function of RF with the points colour-coded by year. The dashed lines in each panel represent a linear fit to the Cycle 23 RF, with slope 0.12 per cent per RFU.}
    \label{fig:sftemporal}
\end{figure*}

For comparison, we show in Figure~\ref{fig:sftemporal} the same analysis for SolarFLAG data. We note that for the pre-1992 data the synthetic data follow the cycle more closely than do the real observations; this is due to the higher noise levels in the observations in the early days of the network. The SolarFLAG data do not reproduce the different sensitivity in the most recent cycle, which provides some reassurance that this is not an artefact of the BiSON duty cycle. 

\subsection{Comparison with results from mode fitting}

\begin{figure*}
\centering

%\includesvg[width=0.9\linewidth]{fig5.svg}%{acomp_22-3-17.svg}
\includegraphics[width=0.9\linewidth]{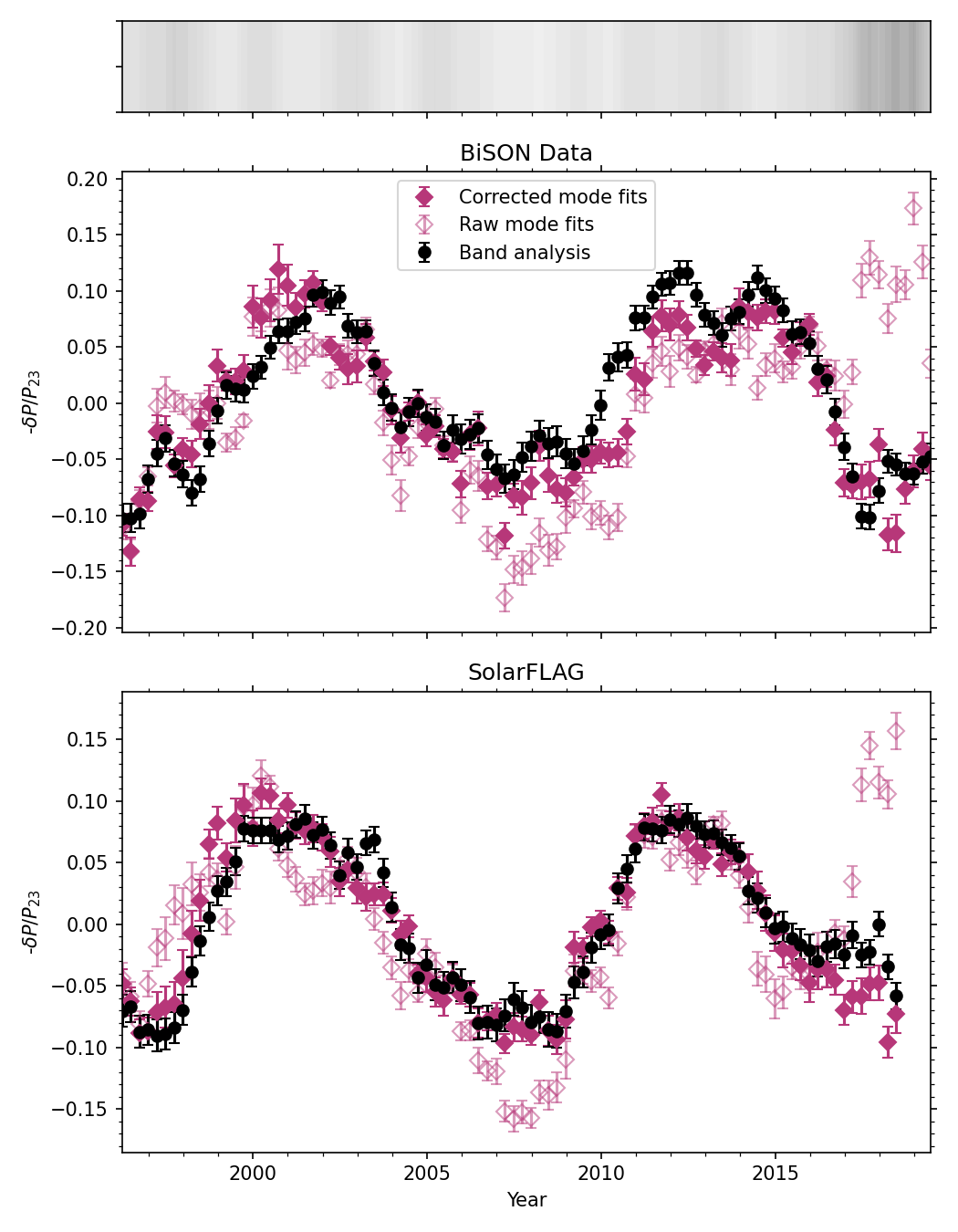}
\caption{Sign-reversed power variation in the five-minute band from spectral band analysis (black circles)  compared with that from MLE fits to modes of $l \leq 2$ in the same frequency range, with (open diamonds) and without (filled diamonds) an empirical correction for the fill, for BiSON (upper panel) and SolarFLAG (lower panel). The grey-scale band at the top indicates the fill, with darker shades indicating lower values; black would correspond to zero fill and white to 100 per cent, while the duty cycle for the data included in the plot spans a range from 56 to 86 per cent.}
\label{fig:fitcomp}
\end{figure*}

In order to check the performance of our method of measuring the mode power changes against the more conventional approach of fitting individual modes, we used maximum likelihood (MLE) fitting of the mode parameters as described by \cite{2015MNRAS.454.4120H} and averaged the changes in the logarithm of the power (amplitude squared) over modes of degree 0, 1, and 2 with frequencies between 2.5 and 4\,mHz, for the data from 1996 onwards, that is, for Cycles 23 and 24. In Figure~\ref{fig:fitcomp} we show the results compared with those for the power in the same frequency range, both for the BiSON observations and SolarFLAG. The agreement is good except where the duty cycle drops, and the comparison with the SolarFLAG result shows that it is the mode fits that are more sensitive to the duty cycle, as is particularly evident after 2017 when the fill dropped. This sensitivity to the fill in the fitted power is to be expected, because of the way that convolution of the mode spectrum with that of the duty cycle redistributes power away from the main peak. For this reason we also show a ``corrected'' version of the result from mode fitting, where the power has been divided by the fill raised to the power 0.72.  
The exponent of 0.72 here was arrived at empirically, by  using a specially constructed synthetic time series with no intrinsic  solar-cycle variation and adjusting the exponent of the duty cycle factor to minimize the slope of power change as a function of fill. This correction brings the power changes from the fitting into reasonable agreement with those from our simple spectral-power analysis. We note that these corrections may not hold when the duty cycle is very low. In summary, this analysis lends us further confidence in the robustness of our results
on the variation of the mode power from the spectral-band analysis.

\section{Discussion and conclusions}

We have analysed the power variations in Sun-as-a-star radial velocity measurements from the BiSON network over four solar cycles by a new method, using the integrated spectral power in selected frequency ranges. This method is robust when the duty cycle is changing, as it avoids the need to correct for power redistributed away from the main peak by window-function effects. We see a clear variation with the solar cycle, as has also been reported in previous work, but the power does not appear to be a simple linear function of the activity level when we consider more than one solar cycle. In particular, the minimum-to-maximum change in mode power is similar in Cycles 23 and 24, in spite of the reduced activity of Cycle 24 compared to Cycle 23 (as seen in the RF). 

The similar scale of amplitude shift in cycles of different strength is in contrast to what it is seen with the mode frequency changes. 
It is well known that the frequencies of the oscillations are changed by the activity. There has been considerable effort to understand what the frequency differences are correlated with. No correlation is found to be perfect over a range of different cycles, although the 10.7\,cm radio flux is generally considered a good proxy. The mean unsigned magnetic-field strength has also been used as a proxy, for example by \cite{2018MNRAS.480L..79H}. They found subtle differences in the sensitivity of the frequencies to activity proxies in different cycles. In contrast, the differences between the activity--power relationship in Cycles 23 and 24 are much less subtle; the reduction in power at solar maximum is similar in scale for the two cycles even though Cycle 24 was weaker in all activity proxies. This could be interpreted as meaning either that the scale of the amplitude variation is roughly the same for all cycles (although the evidence for the first two cycles we observed, Cycles 21 and 22, is not as compelling due to the low duty cycle and noise issues) or that the sensitivity of the mode amplitude to activity is different in different cycles, specifically higher in Cycle 24 than in Cycle 23. In either case, this could point to changes in the outer layers of the Sun, or it could mean that the overall picture is more complex than previously thought.

One possibility worth considering is that different kinds of surface magnetic activity are driving the changes in frequencies and in mode power. To investigate this, we followed the analysis of \cite{2019MNRAS.489L..86C}, who divided the magnetic flux observed at the Wilcox Solar Observatory (WSO)  into ``strong'' ($>15$\,G magnetic field strength in a pixel, considered as a proxy for the field in active regions) and ``weak'' ($<15$\,G, considered as a proxy for the flux in ephemeral regions, plage, etc) components. They found that the frequency variation outside periods of solar minimum was dominated by the strong-flux component, while the weak-flux component was not well correlated with the frequency variations even though it makes up the majority of the flux except around solar maximum. Is it possible that the weak flux plays more of a role in the damping and excitation of the modes? 

\begin{figure*}
\includegraphics[width=\linewidth]{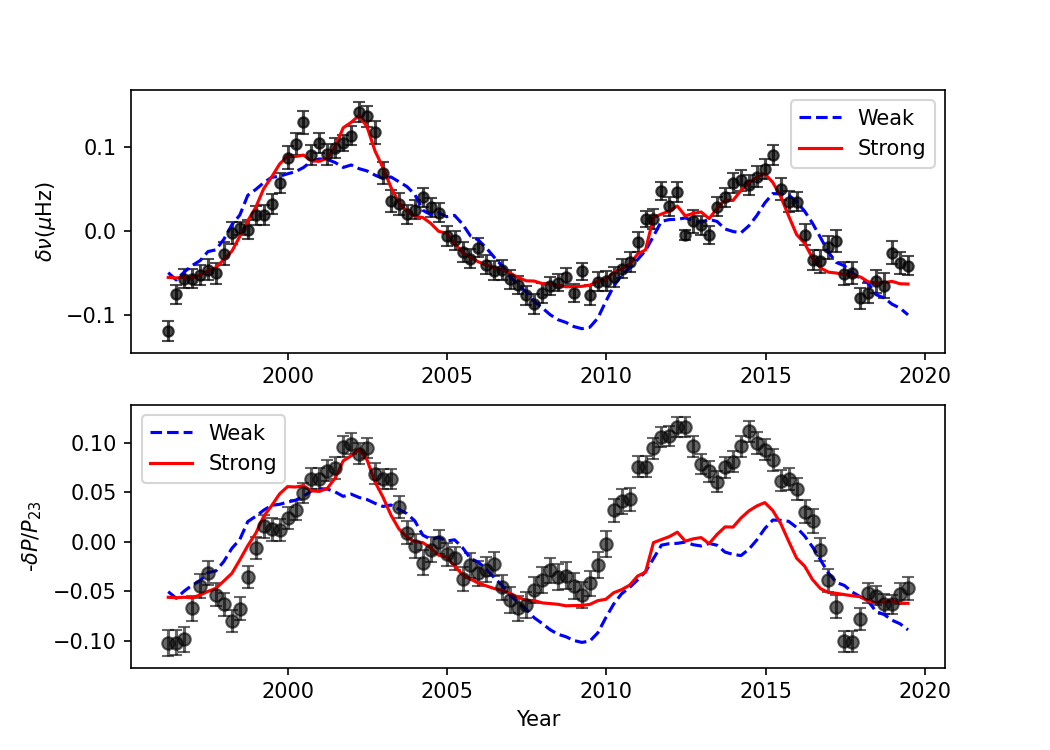}%{splitflux2_22m.png}
\caption{Frequency variations from mode fitting (top) for BiSON observations from Cycles 23 and 24, and the sign-reversed power variation (bottom) from our current analysis. The dashed and solid curves represent separate linear fits to the ``weak'' and ``strong'' components, respectively, of the magnetic flux for Cycle 23.}
\label{fig:yeplot}
\end{figure*}

In Figure~\ref{fig:yeplot} we show both the frequency shifts (obtained using the same mode fits used for Figure~\ref{fig:fitcomp}) and the sign-reversed power variation from our current analysis, with linear fits to the weak and strong magnetic-flux components in Cycle 23. We can see that the difference between the two cycles is less marked in the weak-flux variation, and for example the two peaks of Cycle 24 are almost equal, which is similar to what we see in the power variation, whereas in the strong flux the second peak is stronger (which is reflected in the frequency variation). However, the weak flux still does not predict the Cycle 24 variation well when extrapolated from Cycle 23.

It should be noted that the data fill at the two maxima is high and so this result is independent of the method used to derive the amplitude.  The fact that the amplitudes at the three solar minima covered are very similar for the method we have selected is also noteworthy, but this result is dependent on the method used for the amplitude. 

It is usual \citep[see, for example, ][]{2000MNRAS.313...32C} to interpret the amplitude variation as caused by an underlying change in the mode damping. Although these changes do follow the solar activity cycle, they appear more consistent between cycles than are the surface effects, perhaps suggesting more global influences. This is yet more evidence that not all solar cycles are alike in their effect on the acoustic modes, and it will be interesting to see how the amplitude variation develops
in Cycle 25. It would also be interesting to look directly at the damping via the line width variation in future work.

\section*{Acknowledgements}
We would like to thank all those who have been associated with BiSON over the years.

This research made use of the Python libraries AstroPy \citep[][]{astropy:2013,astropy:2018}, NumPy \citep[][]{harris2020array}, MatPlotLib \citep[][]{Hunter:2007}, and SciPy \citep[][]{2020SciPy-NMeth}, and of NASA’s Astrophysics Data System.

This work was supported by STFC grant ST/V000500/1. Funding for the Stellar Astrophysics Centre (SAC) is provided by The Danish National Research Foundation, grant reference DNRF106. MBN acknowledges support from the UK Space Agency.

We thank the anonymous referee for helpful comments.

\section*{Data Availability}

The BiSON time series analysed here is available at \url{http://bison.ph.bham.ac.uk/opendata}.

The SolarFLAG time series and the weak/strong WSO magnetic-flux data are available on reasonable application to the authors. The original WSO data are available at \url{http://wso.stanford.edu}.

The daily 10.7\,cm flux data can be downloaded from \url{ftp://ftp.seismo.nrcan.gc.ca/spaceweather/solar_flux/daily_flux_values}.



\bibliographystyle{mnras}
\bibliography{bison_p} % if your bibtex file is called example.bib

\begin{thebibliography}{}
\makeatletter
\relax
\def\mn@urlcharsother{\let\do\@makeother \do\$\do\&\do\#\do\^\do\_\do\%\do\~}
\def\mn@doi{\begingroup\mn@urlcharsother \@ifnextchar [ {\mn@doi@}
  {\mn@doi@[]}}
\def\mn@doi@[#1]#2{\def\@tempa{#1}\ifx\@tempa\@empty \href
  {http://dx.doi.org/#2} {doi:#2}\else \href {http://dx.doi.org/#2} {#1}\fi
  \endgroup}
\def\mn@eprint#1#2{\mn@eprint@#1:#2::\@nil}
\def\mn@eprint@arXiv#1{\href {http://arxiv.org/abs/#1} {{\tt arXiv:#1}}}
\def\mn@eprint@dblp#1{\href {http://dblp.uni-trier.de/rec/bibtex/#1.xml}
  {dblp:#1}}
\def\mn@eprint@#1:#2:#3:#4\@nil{\def\@tempa {#1}\def\@tempb {#2}\def\@tempc
  {#3}\ifx \@tempc \@empty \let \@tempc \@tempb \let \@tempb \@tempa \fi \ifx
  \@tempb \@empty \def\@tempb {arXiv}\fi \@ifundefined
  {mn@eprint@\@tempb}{\@tempb:\@tempc}{\expandafter \expandafter \csname
  mn@eprint@\@tempb\endcsname \expandafter{\@tempc}}}

\bibitem[\protect\citeauthoryear{{Astropy Collaboration} et~al.,}{{Astropy
  Collaboration} et~al.}{2013}]{astropy:2013}
{Astropy Collaboration} et~al., 2013, \mn@doi [\aap]
  {10.1051/0004-6361/201322068}, \href
  {http://adsabs.harvard.edu/abs/2013A%26A...558A..33A} {558, A33}

\bibitem[\protect\citeauthoryear{{Astropy Collaboration} et~al.,}{{Astropy
  Collaboration} et~al.}{2018}]{astropy:2018}
{Astropy Collaboration} et~al., 2018, \mn@doi [\aj] {10.3847/1538-3881/aabc4f},
  \href {https://ui.adsabs.harvard.edu/abs/2018AJ....156..123A} {156, 123}

\bibitem[\protect\citeauthoryear{{Broomhall}, {Pugh}  \&
  {Nakariakov}}{{Broomhall} et~al.}{2015}]{2015AdSpR..56.2706B}
{Broomhall} A.~M.,  {Pugh} C.~E.,   {Nakariakov} V.~M.,  2015, \mn@doi
  [Advances in Space Research] {10.1016/j.asr.2015.04.018}, \href
  {https://ui.adsabs.harvard.edu/abs/2015AdSpR..56.2706B} {56, 2706}

\bibitem[\protect\citeauthoryear{{Chaplin} et~al.,}{{Chaplin}
  et~al.}{1996}]{1996SoPh..168....1C}
{Chaplin} W.~J.,  et~al., 1996, \mn@doi [\solphys] {10.1007/BF00145821}, \href
  {http://adsabs.harvard.edu/abs/1996SoPh..168....1C} {168, 1}

\bibitem[\protect\citeauthoryear{{Chaplin}, {Elsworth}, {Isaak}, {Miller}  \&
  {New}}{{Chaplin} et~al.}{2000}]{2000MNRAS.313...32C}
{Chaplin} W.~J.,  {Elsworth} Y.,  {Isaak} G.~R.,  {Miller} B.~A.,   {New} R.,
  2000, \mn@doi [\mnras] {10.1046/j.1365-8711.2000.03176.x}, \href
  {https://ui.adsabs.harvard.edu/abs/2000MNRAS.313...32C} {313, 32}

\bibitem[\protect\citeauthoryear{{Chaplin} et~al.,}{{Chaplin}
  et~al.}{2019}]{2019MNRAS.489L..86C}
{Chaplin} W.~J.,  et~al., 2019, \mn@doi [\mnras] {10.1093/mnrasl/slz132}, \href
  {https://ui.adsabs.harvard.edu/abs/2019MNRAS.489L..86C} {489, L86}

\bibitem[\protect\citeauthoryear{Davies, Chaplin, Elsworth  \& Hale}{Davies
  et~al.}{2014}]{doi:10.1093/mnras/stu803}
Davies G.~R.,  Chaplin W.~J.,  Elsworth Y.~P.,   Hale S.~J.,  2014, \mn@doi
  [Monthly Notices of the Royal Astronomical Society] {10.1093/mnras/stu803},
  441, 3009

\bibitem[\protect\citeauthoryear{{Elsworth}, {Howe}, {Isaak}, {McLeod},
  {Miller}, {Speake}, {Wheeler}  \& {New}}{{Elsworth}
  et~al.}{1993}]{1993MNRAS.265..888E}
{Elsworth} Y.,  {Howe} R.,  {Isaak} G.~R.,  {McLeod} C.~P.,  {Miller} B.~A.,
  {Speake} C.~C.,  {Wheeler} S.~J.,   {New} R.,  1993, \mn@doi [\mnras]
  {10.1093/mnras/265.4.888}, \href
  {https://ui.adsabs.harvard.edu/abs/1993MNRAS.265..888E} {265, 888}

\bibitem[\protect\citeauthoryear{{Hale}, {Howe}, {Chaplin}, {Davies}  \&
  {Elsworth}}{{Hale} et~al.}{2016}]{2016SoPh..291....1H}
{Hale} S.~J.,  {Howe} R.,  {Chaplin} W.~J.,  {Davies} G.~R.,   {Elsworth}
  Y.~P.,  2016, \mn@doi [\solphys] {10.1007/s11207-015-0810-0}, \href
  {https://ui.adsabs.harvard.edu/abs/2016SoPh..291....1H} {291, 1}

\bibitem[\protect\citeauthoryear{Harris et~al.,}{Harris
  et~al.}{2020}]{harris2020array}
Harris C.~R.,  et~al., 2020, \mn@doi [Nature] {10.1038/s41586-020-2649-2}, 585,
  357

\bibitem[\protect\citeauthoryear{{Hindman} \& {Brown}}{{Hindman} \&
  {Brown}}{1998}]{1998ApJ...504.1029H}
{Hindman} B.~W.,  {Brown} T.~M.,  1998, \mn@doi [\apj] {10.1086/306128}, \href
  {https://ui.adsabs.harvard.edu/abs/1998ApJ...504.1029H} {504, 1029}

\bibitem[\protect\citeauthoryear{{Howe}, {Chaplin}, {Elsworth}, {Hill}, {Komm},
  {Isaak}  \& {New}}{{Howe} et~al.}{2003}]{2003ApJ...588.1204H}
{Howe} R.,  {Chaplin} W.~J.,  {Elsworth} Y.~P.,  {Hill} F.,  {Komm} R.,
  {Isaak} G.~R.,   {New} R.,  2003, \mn@doi [\apj] {10.1086/374336}, \href
  {https://ui.adsabs.harvard.edu/abs/2003ApJ...588.1204H} {588, 1204}

\bibitem[\protect\citeauthoryear{{Howe}, {Komm}, {Hill}, {Haber}  \&
  {Hindman}}{{Howe} et~al.}{2004}]{2004ApJ...608..562H}
{Howe} R.,  {Komm} R.~W.,  {Hill} F.,  {Haber} D.~A.,   {Hindman} B.~W.,  2004,
  \mn@doi [\apj] {10.1086/392525}, \href
  {https://ui.adsabs.harvard.edu/abs/2004ApJ...608..562H} {608, 562}

\bibitem[\protect\citeauthoryear{{Howe}, {Davies}, {Chaplin}, {Elsworth}  \&
  {Hale}}{{Howe} et~al.}{2015}]{2015MNRAS.454.4120H}
{Howe} R.,  {Davies} G.~R.,  {Chaplin} W.~J.,  {Elsworth} Y.~P.,   {Hale}
  S.~J.,  2015, \mn@doi [\mnras] {10.1093/mnras/stv2210}, \href
  {https://ui.adsabs.harvard.edu/abs/2015MNRAS.454.4120H} {454, 4120}

\bibitem[\protect\citeauthoryear{{Howe}, {Basu}, {Davies}, {Ball}, {Chaplin},
  {Elsworth}  \& {Komm}}{{Howe} et~al.}{2017}]{2017MNRAS.464.4777H}
{Howe} R.,  {Basu} S.,  {Davies} G.~R.,  {Ball} W.~H.,  {Chaplin} W.~J.,
  {Elsworth} Y.,   {Komm} R.,  2017, \mn@doi [\mnras] {10.1093/mnras/stw2668},
  \href {https://ui.adsabs.harvard.edu/abs/2017MNRAS.464.4777H} {464, 4777}

\bibitem[\protect\citeauthoryear{{Howe}, {Chaplin}, {Davies}, {Elsworth},
  {Basu}  \& {Broomhall}}{{Howe} et~al.}{2018}]{2018MNRAS.480L..79H}
{Howe} R.,  {Chaplin} W.~J.,  {Davies} G.~R.,  {Elsworth} Y.,  {Basu} S.,
  {Broomhall} A.~M.,  2018, \mn@doi [\mnras] {10.1093/mnrasl/sly124}, \href
  {https://ui.adsabs.harvard.edu/abs/2018MNRAS.480L..79H} {480, L79}

\bibitem[\protect\citeauthoryear{Hunter}{Hunter}{2007}]{Hunter:2007}
Hunter J.~D.,  2007, \mn@doi [Computing in Science \& Engineering]
  {10.1109/MCSE.2007.55}, 9, 90

\bibitem[\protect\citeauthoryear{{Rajaguru}, {Basu}  \& {Antia}}{{Rajaguru}
  et~al.}{2001}]{2001ApJ...563..410R}
{Rajaguru} S.~P.,  {Basu} S.,   {Antia} H.~M.,  2001, \mn@doi [\apj]
  {10.1086/323780}, \href
  {http://adsabs.harvard.edu/cgi-bin/nph-bib_query?bibcode=2001ApJ...563..410R&db_key=AST}
  {563, 410}

\bibitem[\protect\citeauthoryear{{Tapping}}{{Tapping}}{2013}]{2013SpWea..11..394T}
{Tapping} K.~F.,  2013, \mn@doi [Space Weather] {10.1002/swe.20064}, \href
  {http://adsabs.harvard.edu/abs/2013SpWea..11..394T} {11, 394}

\bibitem[\protect\citeauthoryear{Virtanen et~al.,}{Virtanen
  et~al.}{2020}]{2020SciPy-NMeth}
Virtanen P.,  et~al., 2020, \mn@doi [Nature Methods]
  {10.1038/s41592-019-0686-2}, \href {https://rdcu.be/b08Wh} {17, 261}

\makeatother
\end{thebibliography}

%\begin{thebibliography}{99}
%\end{thebibliography}

%%%%%%%%%%%%%%%%%%%%%%%%%%%%%%%%%%%%%%%%%%%%%%%%%%

%%%%%%%%%%%%%%%%% APPENDICES %%%%%%%%%%%%%%%%%%%%%

%\appendix



%%%%%%%%%%%%%%%%%%%%%%%%%%%%%%%%%%%%%%%%%%%%%%%%%%

% Don't change these lines
\bsp	% typesetting comment
\label{lastpage}
\end{document}